\documentclass[twocolumn,showpacs,preprintnumbers,amsmath,amssymb]{revtex4-1}
\usepackage{amsfonts}
\usepackage{amsmath}
\usepackage{graphicx}
\usepackage{dcolumn}
\usepackage{bm}
\usepackage{overpic}
\usepackage{booktabs}
\usepackage{color}
\usepackage{threeparttable}

\begin{document}
\title{Control core of undirected complex networks}
\author{Zhengzhong Yuan$^1$}
\author{Jingwen Li$^2$}
\author{Chen Zhao$^3$}\email{tczxz007@163.com}
\author{Li Hu$^3$}
\author{Zhesi Shen$^4$}\email{shenzhs@mail.las.ac.cn}

\affiliation{1, School of Mathematics and Statistics, Minnan Normal University, Zhangzhou, 363000, P. R. China\\
2, Department of Physics, University of Arkansas, Fayetteville, Arkansas, 72701, USA\\
3, College of Computer and Cyber Security, Hebei Normal University, Shijiazhuang, 050024, P. R. China\\
4, National Science Library, Chinese Academy of Sciences, Beijing, 100190, P. R. China\\
}
\date{\today}

\begin{abstract}
With the development of complex networks, many researchers have paid greater attention to studying the control of complex networks over the last decade. Although some theoretical breakthroughs allow us to identify all driver nodes, we still lack an efficient method to identify the driver nodes and understand the roles of individual nodes in contributing to the control of a large complex network. Here, we apply a leaf removal process (LRP) to find a substructure of an undirected network, which is considered as the control core of the original network. Based on a strict mathematical proof, the control core obtained by the LRP has the same controllability as the original network, and it contains at least one set of driver nodes. With this method, we systematically investigate the structural property of the control core with respect to different average degrees of the original networks ($\langle k \rangle$). We denote the node density ($n_\text{core}$) and link density ($l_\text{core}$) to characterize the control core when applying the LRP, and we study the impact of $\langle k \rangle$ on $n_\text{core}$ and $l_\text{core}$ in two artificial networks: undirected Erd\"os-R\'enyi (ER) random networks and undirected scale-free (SF) networks. We find that $n_\text{core}$ and $l_\text{core}$ both change nonmonotonously with increasing $\langle k \rangle$ in the two typical undirected networks. With the aid of core percolation theory, we can offer the theoretical predictions for both $n_\text{core}$ and $l_\text{core}$ as a function of $\langle k \rangle$. Then, we recognize that finding the driver nodes in the control core is much more efficient than in the original network by comparing $n_{\text{D}}$, the controllability of the original network, and $n_{\text{core}}$, regardless of how $\langle k \rangle$ increases.
Finally, we consider some empirical networks, finding that more than $20\%$ nodes can be removed by the LRP, and the networks can be divided into many smaller components, which greatly improves the efficiency of identifying and analyzing the driver nodes of the empirical networks. These findings deepen our understanding of the relationship between structure and controllability of undirected networks, and provide an effective method for a more detailed study on the driver nodes in large undirected complex networks.

\end{abstract}
\keywords{}
\pacs{}
\maketitle

\section{Introduction}
Controlling complex systems is an important research topic of complexity science, and should rely on our understanding of the principles in the extraordinarily complex system of nature.
Past decades have witnessed the rich development of complex science, especially in the study of topological analysis and kinetic mechanisms~\cite{Newman:2003,BLMCH:2006}. Since the discovery of scale-free networks~\cite{BA:1999}, there have been numerous theoretical and empirical studies on the architecture of various complex networks~\cite{JMBO:2001,AB:2002}.
As these studies deepen our understanding of structural mechanisms and dynamic mechanisms of complex networks~\cite{SBGC:2007,RME:2009}, researchers are beginning to pay greater attention to the problem of controlling complex networks.
Ref.~\cite{Liu2011} was published as a pioneering research work on the control of complex networks. The authors creatively established a theoretical framework to analyze the controllability of large-scale complex networks using linear time-invariant (LTI) system theory, called structural controllability theory (SCT). Based on maximum matching theory of digraphs, it has been theoretically proven that the minimum number of driver nodes (denoted as $N_\text{D}$) required to be controlled by imposing independent signals to fully control the system is equal to the number of unmatched nodes. The fraction of driver nodes (denoted as $n_\text{D}$) usually measures the controllability of a network.
SCT not only is an extension of conventional control problems in complex networks but also provides a new method and insight for investigating the control problem of complex networks. However, SCT is valid on systems with a structural coupling matrix, which means that it can only analyze the systems with all directed links represented by independent free parameters.
There are actually many coupled systems with fixed-weight or unweighted correlations. For these systems, Yuan et al. proposed exact controllability theory (ECT) to address the control problem in more general situations~\cite{Yuan2013}. ECT claims that $N_\text{D}$ can be obtained by calculating the maximum geometric multiplicity of the coupling matrix. Based on a strict mathematical proof from the Popov-Belevitch-Hautus (PBH) principle, ECT can be applied to arbitrary networks and link weights: directed or undirected and weighted or unweighted. Inspired by these two studies, many researchers have moved their focus toward the problem of controlling complex networks~\cite{LB:2016}, including the improvement of controllability theory~\cite{YZD:2013,RR:2014,NV:2012}, optimization of controllability~\cite{WNL:2012,XLHB:2014}, controllability of dynamical systems~\cite{ZWL:2015,ZZW:2017}, target control~\cite{GLDB:2014}, control node analysis~\cite{JLC:2013,IGS:2015}, and so on.

Although SCT and ECT allow us to identify all driver nodes, it is still computationally difficult to analyze the controllability of a large complex network, that contains thousands of nodes. Based on SCT, some researchers have attempted to build the connection between core percolation and controllability to analyze the functions of the nodes in the core structure for directed uncorrelated random networks~\cite{JP:2014}. However, we still lack an efficient method to identify the driver nodes and understand the roles of individual nodes in contributing to controllability in large undirected networks, weighted or unweighted, with arbitrary degree distribution.
In this paper, we apply a leaf removal process (LRP) to find the control core, which contains at least one set of driver nodes of the original network and has the same controllability as the original network. The LRP in our work removes the leaf, which is defined as a one-degree node together with its neighbor, and retains all of the isolated nodes. According to ECT, it is rigorously proven that the LRP can theoretically ensure invariant controllability of each removal step.
Applying our framework in undirected Erd\"os-R\'enyi (ER) random networks and undirected scale-free (SF) networks, we systematically study the node density (denoted as $n_\text{core}$) and the link density (denoted as $l_\text{core}$) of the control core with different average degrees of the original networks ($\langle k \rangle$).
It is found that $n_{\text{core}}$ and $l_\text{core}$ of the control core change nonmonotonously with increasing $\langle k \rangle$. As $\langle k \rangle$ increases, $n_{\text{core}}$ decreases and $l_\text{core}$ remains zero until the onset point. When $\langle k \rangle$ exceeds the value of the onset point, $n_{\text{core}}$ and $l_\text{core}$ both begin to increase.
Based on core percolation theory, analytical predictions can be obtained to evaluate $n_{\text{core}}$ and $l_{\text{core}}$ as functions of $\langle k \rangle$ for ER random networks and scale-free networks, respectively.
We then compare $n_{\text{D}}$ and $n_{\text{core}}$, and learn that identifying and analyzing the driver nodes of the original network in its control core is actually more efficient than in the original network, regardless of how $\langle k \rangle$ increases.
Finally, we consider some empirical networks, finding that more than $20\%$ of nodes in empirical networks can be removed, and many isolated nodes are retained in the control core, which are the exact driver nodes of the original empirical network. Although some connected components are retained in the control core, we find that the largest connected component is still much smaller than the original network. This indicates that finding one set of driver nodes in the control core is also a more efficient method for empirical networks.

\section{Results}
\subsection{Control core of undirected networks}
Consider a networked linear system with fixed structure of $N$ nodes, which is described by the following linearly ordinary differential equation:
\begin{equation}
\label{eq1}
\dot{\text{\bf X}} = A{\text{\bf X}}+B{\text{\bf u}},
\end{equation}
where vector ${\bf X} = (x_ {1}, \cdots, x_ {N})^{\text{T}} $ represents the state of all nodes, and $A = (a_ {ij}) _ {N \times N} $ is the coupling matrix of the network, with $a_ {ij}$ being the direct link weight of node $j$ to node $i$ (for undirected networks, $a_{ij}=a_{ji}$). $u $ is the controller with ${\bf u} = (u_ {1}, u_ {2}, \cdots, u_m) ^ {\text{T}} $, and $B $ is a $N \times m $ input matrix of the system, where $b_{ij}$ is the strength of controller $u_j$ on node $i$. For such a system, when the coupling matrix $A$ is a sparse matrix, where its zero eigenvalue is dominant, by the exact controllability theory (ECT)~\cite{Yuan2013}, the minimum number of controller $N_{\text{D}}$ of the network can be obtained by the following
\begin{equation}
\label{eq2}
N_{\text{D}} = N-\text{rank}(A);
\end{equation}
All of the nodes, which are independently controlled by controller $u_i~(i=1,2,\cdots,N_\text{D})$, are called driver nodes ~\cite{Liu2011,Yuan2013}. According to eq.~(\ref{eq2}), the driver nodes can be considered as the nodes corresponding to linearly dependent rows of coupling matrix $A$, which can be obtained by performing elementary column transformation on matrix $A$~\cite{Yuan2013}. In other words, if we remove one node corresponding to a linearly independent row of $A$ and its links with other nodes, the revised network retains the same controllability as the original network.
For this reason, we employed a leaf removal process (LRP) to remove the nodes and to reduce the structure of an arbitrary undirected network. The LRP is defined as removing a leaf, where a leaf is composed of a one-degree node and its neighbor with their links, one by one, until no leaf exists, following the principle in Ref.~\cite{core_percolation1}. However, all the isolated nodes are retained in the revised network, which is quite different from the definition of the core in many studies~\cite{core_percolation1, core_percolation2, core_percolation3}. In each step of the LRP, we randomly pick one leaf as the removal leaf and remove it from the network. We then update the node degrees and continue the former operation on the revised network until leaves no longer exist in the final subnetwork. The LRP can be theoretically proven without changing the linearly dependent part of the coupling matrix~\cite{Yuan2013}. Thus, we can consider the final subnetwork gained by the LRP as the control core of the original network. The control core has the same controllability as the original network, and it contains at least one set of driver nodes of the original network. It is emphasized that we cannot find all of the combinations of driver nodes of the original network in the control core, but we can find at least one set of driver nodes of the original network.

\subsection{Matrix representation of LRP}
Without loss of generality, we consider an undirected network with a node of one degree marked as Node~$1$ and its neighbor marked as Node~$2$. The coupling structure of Node~$1$ and Node~$2$ is a leaf, and the coupling matrix of the whole undirected network is
\begin{equation}
A = \left[
\begin{array}{ccc}
0&1&0\\
1&0&\alpha^\text{T}\\
0&\alpha &A_0
\end{array}
\right],
\end{equation}
where $\alpha$ denotes other neighbors of Node~$2$, $A_0$ is the coupling matrix of the other nodes, and the four $0$s represent the vectors with appropriate dimensions. Then, we apply the following elementary transformation on A for describing the LRP: multiplying the first block column by $-\alpha^\text{T}$ and adding the new column to the third block column, which is analogous to modifying the third block row by adding the first block row multiplied by $-\alpha^\text{T}$. We obtain
\begin{equation}
A_1 = \left[
\begin{array}{ccc}
0&1&0\\
1&0&0\\
0&0&A_0
\end{array}
\right],
\end{equation}
Since the first row (column) and second row (column) are independent of the other rows (columns), we can delete the Node~$1$ and Node~$2$ corresponding to the first two rows and columns without changing the linearly dependent lines. Note that the remaining network $A_0$ is intact, such that the links among nodes belonging to $A_0$, including the driver nodes, are unchanged. According to ECT, the driver nodes can be obtained from matrix $A_0$ instead of matrix $A$ as their same linearly dependent lines after a removal step. This means that the revised network after one step of the LRP has the same controllability as the original network. We repeat this procedure until there is no matrix block for the leaf. All of the lines filled with zeros, which correspond to the isolated nodes, cannot be removed during the LRP, and these lines are retained in the final revised matrix. The final revised matrix, which has the same linearly dependent lines as the original matrix $A$, actually corresponds to the coupling matrix of the control core of the original network.
It should be noted that the above certification process is only based on topological information of the network, regardless of the link weights. Thus, this framework, using the LRP to find the control core, is applicable to unweighted or weighted networks.

\begin{figure}[t!]
\begin{center}
\includegraphics[width=\linewidth]{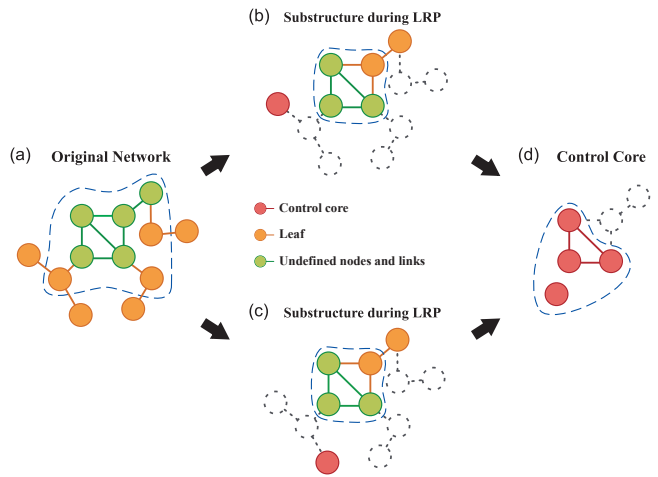}
\caption{{\bf Illustration of the leaf removal process}. All nodes in networks are marked in three colors for a current removal step. The yellow node represents the leaf of the current removal step. The red node is determined to be a part of the control core. The green node represents the undefined nodes, as it does not belong to any leaf or the control core at the current step of the LRP. The substructures ({\bf b, c}) are both obtained from the original structure ({\bf a}) after several steps of the LRP, which is determined by the order of removed leaves, but the number of nodes and the number of links remaining in the control core are constant ({\bf d}).}
\label{fig1}
\end{center}
\end{figure}

In Fig.~\ref{fig1}, we illustrate the LRP on a sparse undirected network with $12$ nodes. All nodes and links are divided into three categories with different colors: the nodes and links remaining in the control core are marked as red, the leaves are marked as yellow, and the undefined nodes and links are marked as green. At the beginning step of the LRP, we first find all leaves of the original network, which are marked as yellow as shown in Fig.~\ref{fig1}(a). We randomly remove one of the leaves, which is composed of a one-degree node and its neighbor. After each step of the LRP, some undefined green nodes convert to new leaves due to updating the degrees of the retained nodes, and all of the new one-degree nodes with their neighbors build the alternative set of removal leaves for the next step. When no leaf exists, the final connected component and all isolated nodes construct the control core of the original network, as shown in Fig.~\ref{fig1}(d).

It is worth noting that if a node only belongs to a single leaf, we can directly remove it without considering the removal order during the LRP. However, if a node belongs to multiple leaves, it is obvious that a different removal order may cause a different revised substructure, as displayed in Fig.~\ref{fig1}(b) and Fig.~\ref{fig1}(c).
Although the different removal orders lead to different components of the control core, the number of nodes and the number of links ultimately remaining in the control core are constant, which indicates that the core is unique despite the different labeled nodes. In other words, a different removal order is simply equivalent to picking different combinations of driver nodes, which is the same function as performing elementary transformation on the coupling matrix of the original network based on ECT.

\subsection{Structural property of the control core}
According to the above proof and analysis, we conclude that the control core of a sparse undirected network can be obtained by performing the LRP on the original network. To understand the property of the control core with respect to different link densities, we apply the LRP on two typical undirected artificial networks: undirected ER random networks and undirected SF networks.
We use the node density $n_\text{core}$ and the link density $l_\text{core}$ to sketch the structural property of the control core. We also define that $n_\text{core} = N_\text{core}/N$ and $l_\text{core} = L_\text{core}/L$, where $N$ is denoted as the size of the original network, $L$ is denoted as the number of links in the original network. $N_\text{core}$ and $L_\text{core}$ are denoted as the number of nodes and links remaining in the control core, respectively.

\begin{figure}[t!]
\begin{center}
\includegraphics[width=\linewidth]{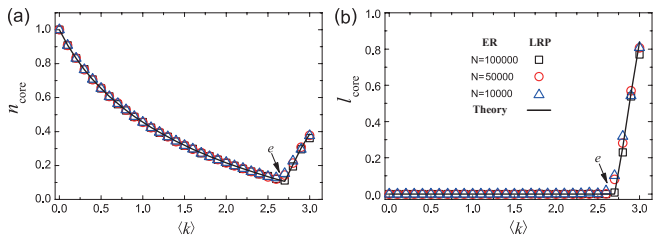}
\caption{{\bf The node density $n_\text{core}$ and the link density $l_\text{core}$ of the control core
as a function of the average degree $\langle k \rangle$ for undirected ER random networks}.
$n_\text{core}$ as a function of $\langle k \rangle$ ({\bf a}) and $l_\text{core}$ as a function of $\langle k \rangle$ ({\bf b}), with three different network sizes: $N = 10000$, $N = 50000$ and $N = 100000$. The colored symbols represent the simulated results from the networks with the LRP applied.
The solid line represents the theoretical predictions from eq.~(\ref{eq:ncore}) and eq.~(\ref{eq:lcore}) based on the expected degree distribution in the $N \rightarrow \infty$. The onset point for undirected ER random networks is proven at $\langle k \rangle = e$ by the numerical solution of eq.~(\ref{eq:w1w2}).
All of the simulations are averaged over $50$ independent calculations.}
\label{fig2}
\end{center}
\end{figure}

Fig.~\ref{fig2} shows $n_\text{core}$ and $l_\text{core}$ as functions of average degree $\langle k \rangle$ in undirected ER random networks.  We see that $n_\text{core}$ is a nonmonotonic function of $\langle k \rangle$. Responding to the increase in $\langle k \rangle$, $n_\text{core}$ decreases first before increasing when $\langle k \rangle$ is greater than the onset point.
For $l_\text{core}$, it remains zero when the value of $\langle k \rangle$ is much lower than the onset point, which suggests that the nodes in the control core are barely connected and that all of them are isolated driver nodes based on ECT. As $\langle k \rangle$ continuously increases until $\langle k \rangle$ exceeds the value of the onset point, the links in the control core begin to appear.

The node density and the link density of the control core can be analytically predicted by using the generating functions based on core percolation theory introduced in Ref.~\cite{core_percolation1,core_percolation2} for undirected ER random networks. In the thermodynamic limit of $N$, node density $n_\text{core}$ becomes
\begin{equation}
n_\text{core} = \frac{2W_1 + W_2^2}{\langle k \rangle}-1,
\label{eq:ncore}
\end{equation}
link density $l_\text{core}$ can be obtained as
\begin{equation}
l_\text{core} = \frac{(W_1-W_2)^2}{2\langle k \rangle},
\label{eq:lcore}
\end{equation}
where $W_1$ and $W_2$ satisfy the following coupling equations:
\begin{equation}
W_1e^{W_2} = \langle k \rangle, \qquad  W_2e^{W_1}=\langle k \rangle,
\label{eq:w1w2}
\end{equation}
and by solving coupling eq.~(\ref{eq:w1w2}), we can obtain the values of $W_1$ and $W_2$ based on the degree distribution. By substituting $W_1$ and $W_2$ into eq.~(\ref{eq:ncore}) and eq.~(\ref{eq:lcore}), we finally obtain the strict solutions of $n_{\text{core}}$ and $l_\text{core}$, respectively.
The onset point of the control core is numerically solved as $\langle k \rangle = e$. All of the solutions are strictly proven when assuming that the network scale is infinite.
The analytical predictions of $n_\text{core}$ and $l_\text{core}$ are both in good agreement with the simulation results, as shown in Fig.~\ref{fig2}.
Note that for an ideal infinity network, as $\langle k \rangle \leq e$, the above coupling eq.~(\ref{eq:w1w2}) has only one solution $W_1 = W_2 = W$, which shows that $l_\text{core}$ should be zero.
In practice, for a finite network with average degree $\langle k \rangle \leq e$, the link density of the control core satisfies $l_\text{core} \approx 0$. When $\langle k \rangle > e$, eq.~(\ref{eq:w1w2}) has two different solutions, $W_1$ and $W_2$, which leads to the edge density $l_\text{core}$ not equaling zero.

\begin{figure}[t!]
\begin{center}
\includegraphics[width=\linewidth]{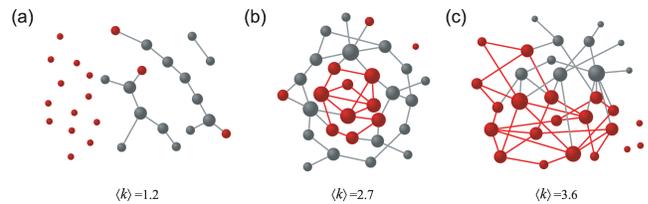}
\caption{{\bf Illustration of the control core in undirected ER random networks with different average node
degrees}. Nodes and links in red belong to the control core, and the other nodes and links in gray are removed
from the original network based on the LRP. The control core contains at least one set of drivers
and their links among them. For $\langle k \rangle=1.2$ ({\bf a}),
all nodes left in the control core are isolated without any links among
them after removing all leaves. For $\langle k \rangle =2.7$ ({\bf b}), the control core of some
connected nodes emerges. When $\langle k \rangle = 3.6$ ({\bf c}), the control core becomes larger
and contains the majority of nodes. Note that in the control core for any
values of $\langle k \rangle$, there exist either connected or isolated nodes,
and the leaves are absent. The size of a node is proportional to its
degree. The network size $N$ is 30.}
\label{fig3}
\end{center}
\end{figure}

A representative control core is illustrated in Fig.~\ref{fig3} for different average degrees of undirected ER random networks. For small average degrees, e.g., 1.2, the size of the control core is relatively large, and there are no links. In fact, a large fraction of nodes in the control core are isolated, and all isolated nodes in the control core are driver nodes based on ECT. When $\langle k \rangle$ approaches $e$, e.g., 2.7, some connected components in the control core emerges, but the size of the control core decreases. There exist some links among the nodes in the control core, as marked by red lines. If $\langle k \rangle$ is larger than $e$, e.g., 3.6, the control core contains a majority of the nodes of the original network. In this case, isolated nodes only represent a small fraction of the control core. The configuration of the control core is determined by the order of leaf removal, but the size of the control core is guaranteed by the core percolation theory.

\begin{figure}[t!]
\begin{center}
\includegraphics[width=\linewidth]{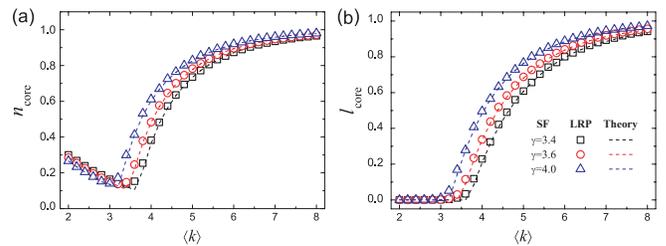}
\caption{{\bf The node density $n_\text{core}$ and link density $l_\text{core}$ as functions of the average degree $\langle k \rangle$ for undirected SF networks}.
The colored hollow polygons represent the simulated results of undirected SF networks with different $\gamma$ values. The colored dashed lines represent the theoretical analytical predictions corresponding to the simulation results with the same $\gamma$, which is proven to be strict, as the size of networks satisfies $N \rightarrow \infty$. The theoretical predictions of $n_\text{core}$ ({\bf a}) and $l_\text{core}$ ({\bf b}) are obtained by the following eq.~(\ref{eq:s2}) and eq.~(\ref{eq:s3}), respectively. The size of simulated undirected SF networks is $N = 3000$, and all calculations are averaged over $500$ iterations.}
\label{fig4}
\end{center}
\end{figure}

We also consider the case of undirected SF networks. The undirected networks are generated following the static growth rule based on the GKK model~\cite{GKK2001}. In this case, the power-law degree distribution can be written as $P(k) \sim k ^ {-\gamma}$.
For a fixed power exponent $\gamma$, we perform LRP on the undirected SF networks for analyzing the impact of average degree on the structural properties of the control core.
The node density $n_\text{core}$ and link density $l_\text{core}$ as functions of average degree $\langle k \rangle$ are shown in Fig.~\ref{fig4}.
Compared with the curves of $n_\text{core}$ and $l_\text{core}$ of the undirected ER random network in Fig.~\ref{fig2}, we obviously find that $\langle k \rangle$ has a similar influence on $n_\text{core}$ and $l_\text{core}$ in undirected SF networks. $n_\text{core}$ is also a nonmonotonic function of $\langle k \rangle$ with the minimum value at the onset point, as shown in Fig.~\ref{fig4}(a), and $l_\text{core}$ remains at zero until the value of $\langle k \rangle$ is greater than the onset point, as shown in Fig.~\ref{fig4}(b). The control core is mainly composed of isolated nodes before the onset point, and the connected components in the control core begin to appear after the onset point.

For different $\gamma$ values, the onset point of the control core differs. The value of the onset point moves toward smaller $\langle k \rangle$ as $\gamma$ increases, which is different from ER random networks.
The value of $\gamma$ represents the dispersion of node degree in undirected SF networks. Smaller $\gamma$ means that the degree distribution of the network decreases slowly, and the degrees of the nodes are more discrete. In other words, the smaller $\gamma$ is, the higher the probability that a large-degree node is the neighbor of a one-degree node, implying that the large-degree nodes are more easily removed by the LRP in a network with smaller $\gamma$, which results in more isolated nodes remaining in the control core responding to the same average degree.

To obtain the theoretical predictions of $n_\text{core}$ and $l_\text{core}$ for undirected SF networks with different exponents, we introduce the rate equation for describing the process of repeatedly applying the LRP on the undirected SF networks~\cite{Dor2013}. Considering an undirected SF network composed of $N$ nodes and $L$ links with degree distribution $P(k)$, the change in the average number of nodes of degree $k$ after one step of the LRP is captured by the following equation
\begin{equation}
\begin{split}
N(k, t+\Delta t) &= N(k, t)-\delta_{k, 1}-\frac{kP(k, t)}{\langle k \rangle_{t}}\\
&+ \frac{\langle k(k-1) \rangle_{t}}{\langle k \rangle_{t}}\frac{(k+1)P(k+1, t)-kP(k, t)}{\langle k \rangle_{t}},
\end{split}
\label{eq:s1}
\end{equation}
where $\Delta t = 1/N$ is a rescaled time for one step of LRP and $N(k,t)$ represents the average number of nodes of degree $k$ at time $t$.
For an undirected, uncorrelated network, given the initial degree distribution, this iterative eq.~(\ref{eq:s1}) can be solved based on the expected degree distribution. The LRP is completed at time $t^*$, where $t^*$ represents the rescaled time via the LRP when there are no longer any existing leaves. The node density $n_\text{core}$ follows
\begin{equation}
n_\text{core} = n(t^*) = \frac{N(1-2t^*)}{N},
\label{eq:s2}
\end{equation}
and the link density $l_\text{core}$ is
\begin{equation}
l_\text{core} = \frac{L(t^*)}{L},
\label{eq:s3}
\end{equation}
where $t^*$ can be obtained by solving eq.~(\ref{eq:s1}) numerically as $P(1, t^*) = 0$. $N(1-2t^*)$ and $L(t^*)$ represent the number of nodes and the number of links at time $t^*$ respectively. The colored dashed lines in Fig.~\ref{fig4} show the theoretical predictions of $n_\text{core}$ and $l_\text{core}$ for undirected SF networks, where cases of different exponents associated with the power-law degree distribution are analyzed. It can be observed that our theoretical predictions are all in good agreement with the results from numerical simulations for different values of $\gamma$.

\subsection{More efficient analysis in the control core}

\begin{figure}[t!]
\begin{center}
\includegraphics[width=\linewidth]{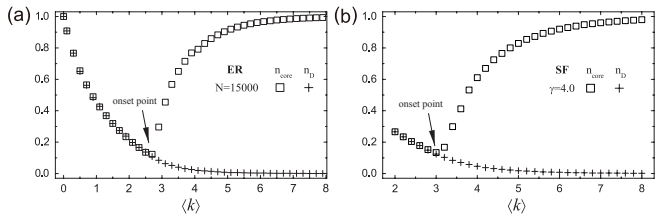}
\caption{{\bf Correspondence between $n_\text{D}$ and $n_\text{core}$ in undirected networks.} The controllability $n_\text{D}$ and the node density $n_\text{core}$ as a function of average degree for undirected ER random networks ({\bf a}) and undirected SF networks ({\bf b}). The size of the simulated undirected ER random network is $N = 15000$, and the size of the simulated undirected SF network is $N = 3000$ with $\gamma = 4$. The square symbols are the fractions of nodes remaining in the control core after the LRP, and the cross symbols are the fractions of the driver nodes obtained by ECT. The data points of undirected ER random networks are the result of averaging $20$ independent realizations. The data points of undirected SF networks are the result of averaging $50$ independent realizations.
}
\label{fig5}
\end{center}
\end{figure}

We also determine the association between $n_\text{core}$ and $n_\text{D}$ in undirected ER random networks and undirected SF networks, as shown in Fig.~\ref{fig5}.
It can be observed that $n_\text{core}$ is equal to $n_\text{D}$ before the onset point in both typical undirected networks, as the control core is composed of isolated nodes without any immediate links. The set of all isolated nodes remaining in the control core is just one set of driver nodes of the original network based on ECT. As $\langle k \rangle$ increases, $n_\text{core}$ is no longer equal to $n_\text{D}$, while $n_\text{D}$ continues to decline and $n_\text{core}$ begins to increase. When $\langle k \rangle$ is greater than the onset point and $n_\text{D}$ of the original network is still greater than $1/N$, the control core gained by the LRP is composed of some isolated nodes and connected components. In this case, one set of driver nodes can be identified as the union of all isolated nodes and the nodes obtained by performing elementary transformation on the coupling matrix of each connected component based on ECT, respectively. The LRP can also reduce the computational complexity of finding one set of driver nodes, because there are much fewer nodes in each component than in the original network. In fact, if the average degree $\langle k \rangle$ is much greater than the onset point and $n_\text{D}$ is equal to $1/N$, we can directly perform the elementary transformation of the original network to find a set of driver nodes, because almost no nodes can be removed by the LRP. In summary, regardless of how many nodes and links remain in the control core, we can recognize that analyzing the controllability of the original network in the control core is a more efficient way to find one set of driver nodes and to more clearly understand the contribution of individual nodes to the controllability of the original network.
\subsection{Empirical networks}

Empirical networks always contain a large number of nodes and links, which makes it difficult to find a set of driver nodes directly by elementary transformation. Fortunately, the majority of the empirical networks are sparse, and the coupling matrix of the networks is always dominated by zero eigenvalues, satisfying the condition for finding the control core through the LRP. We apply the LRP on some undirected networks, which are unweighted or weighted networks from the real world, and we find the control core of each empirical network.

\begin{table}[htbp]
\caption{{\bf Control core of undirected empirical networks.} For each empirical network, the network name and the number of nodes ($N$) are shown. The controllability $n_\text{D}$ is measured by ECT. The quantities $n_\text{core}$ and $l_\text{core}$ are the fractions of remaining nodes and links after the LRP. $n_\text{c}$ represents the ratio of the size of the largest connected component after the LRP to $N$, and $n_\text{i}$ represents the ratio of the number of isolated nodes retained in the control core to $N$.}
\begin{center}
\begin{tabular}{c c c c c c c c c}
\hline
\hline
Index & Name & $N$ & $n_\text{D}$ & $n_\text{core}$ & $l_\text{core}$ & $n_\text{c}$ & $n_\text{i}$ \\
\hline		
$1$ & USA top-500\cite{data-air} & 500 & 0.25 & 0.508 & 0.083 & 0.106 & 0.248\\
$2$ & Internet1997\cite{data-int} & 3015 & 0.625 & 0.625 & 0.0 & 0.0 & 0.625\\
$3$ & Internet2001\cite{data-int} & 10515 & 0.699 & 0.7 & 0.001 & 0.001 & 0.698\\
$4$ & Oregon1\cite{data-kdd2005} & 11174 & 0.703 & 0.704 & 0.001 & 0.001 & 0.702\\
$5$ & Oregon2\cite{data-kdd2005} & 11461 & 0.666 & 0.671 & 0.007 & 0.004 & 0.663\\
$6$ & Email-Enron\cite{data-ema} & 36692 & 0.32 & 0.704 & 0.21 & 0.232 & 0.315\\
\hline
\hline
\end{tabular}
\label{tab1}
\begin{tablenotes}
\footnotesize
 \item $^*$  All of the empirical networks are sparse and exactly zero-eigenvalue dominant. We simply use the topology of each empirical network to confirm the high efficiency of LRP in analyzing the controllability of undirected networks.
\end{tablenotes}
\end{center}
\end{table}

Table~\ref{tab1} displays the results of applying the LRP on a variety of undirected empirical networks. The results for $n_\text{D}$, $n_\text{core}$, and $l_\text{core}$ are listed.
We find that all of the empirical networks could be reduced by performing the LRP, and more than $20\%$ of nodes can be removed from the empirical networks.
Another finding is that for most networks, large numbers of links are removed, and the nodes in the control core are almost isolated, which is obviously concluded from the small $l_\text{core}$. This result suggests that most links have no effect on the controllability of the original networks and that driver nodes of these empirical networks can be identified in an extremely efficient manner through application of the LRP, such as Internet1997.

However, for most of the empirical networks, there are still some nodes and links remaining in the control core. In these cases, we need another strategy to find other driver nodes, except the isolated nodes, by separately applying elementary transformation to each connected component in the control core. To explain the high efficiency of the LRP, we analyze $n_\text{c} = N_\text{c} / N$ and $n_\text{i} = N_\text{i} / N$, where $N_\text{c}$ is the size of the largest connected component and $N_\text{i}$ is the number of isolated nodes after the LRP. $n_\text{i}$ is nearly equal to $n_\text{D}$ for most of the empirical networks, which indicates that most of the driver nodes, the retained isolated nodes, can be found directly in the control core. The rest of driver nodes should be found by performing the elementary transformation on each connected component, respectively. Fortunately, it is shown in Table~\ref{tab1} that $n_\text{c}$ is quite small in all of the empirical networks, which means that the largest connected component in the control core is much smaller than the original network. We know that the computational complexity of performing elementary transformation is $\mathcal O(N^2(logN)^2)$. Thus, the computational complexity of finding one set of driver nodes in the control core is also much smaller than in the original network. For some special networks, such as USA top-500 and Email-Enron, relatively large numbers of nodes and links cannot be removed by the LRP, but each original network has been divided into many smaller connected components based on the small $n_\text{c}$, which still greatly improves the efficiency of finding one set of driver nodes in the control core.

\section{Discussion}

Two existing theoretical frameworks can be used to quantify the controllability of a complex network: structural controllability theory (SCT) and exact controllability theory (ECT). The structural controllability of any directed network is determined by the maximum matching of the network topology based on SCT. Thus, SCT is applicable only to directed networks with arbitrary link weights. The exact controllability is determined by the maximum multiplicity of eigenvalues of the coupling matrix, and it is applicable to any network: directed or undirected and weighted or unweighted.
For directed networks, it is much easier to find the driver nodes and analyze the roles of individual nodes in contributing to the network controllability by SCT~\cite{Liu2011,JP:2014}.
For undirected networks, however, we can only identify the driver nodes by ECT, which is very difficult to do by performing the elementary transformation on a large network.

The framework of the LRP introduced in this paper aims to solve this problem. In particular, the LRP is applied to reduce the size of an undirected network by removing leaves, one after another, to obtain the control core based only on local nodal structural information.
The procedure of the LRP has been theoretically proven to keep the undirected network controllability invariant, which is applicable to weighted or unweighted networks.
Due to unchanged controllability, we can always find at least one set of driver nodes in the control core to bring the original network under full control.
Under this framework, we systematically study $n_\text{core}$ and $l_\text{core}$, which capture the structural property of the control core, with respect to different $\langle k \rangle$ values in undirected ER random networks and undirected SF networks.
We find that $n_\text{core}$ and $l_\text{core}$ are both nonmonotonic with $\langle k \rangle$ in these two typical networks. Inspired by core-percolation theory, we provide analytical predictions to illustrate the nonmonotonic phenomenon of both $n_\text{core}$ and $l_\text{core}$.
Then, we learn that analyzing the controllability of the original network in its control core is much more efficient by comparing $n_\text{D}$ and $n_\text{core}$. Finally, we apply the LRP on some undirected empirical networks and find that most nodes and links of the original empirical network can be removed by the LRP. Obviously, not only in artificial networks but also in empirical networks, the control core gained by the LRP leads to a highly computationally efficient scheme to analyze the controllability of a large undirected network.
At present, we are still incapable of establishing a unified approach to reduce the structure, both in undirected and directed networks based on ECT, because it is very difficult to remove a large number of nodes while maintaining correspondence between the control core and the original network. Nevertheless, the framework developed in this paper still provides an effective and efficient way to identify the driver nodes and understand the role of each node contributing to the controllability for arbitrary undirected networks. We hope that our work can stimulate further efforts toward developing a general method for this purpose and that it will be exploited as an efficient centrality measure for analyzing controllability and designing input matrix in related regulatory networks.

\section*{Acknowledgments}
We thank Professor Wen-Xu Wang for valuable suggestions. This work is supported by the National Natural Science Foundation of China (Nos. 61703136, 61403181, 71974017 and 61672206), the Natural Science Foundation of Fujian (No. 2018J01550), the Natural Science Foundation of Hebei (No. F2017205064), the Natural Science Foundation of Hebei Education Department (No. QN2017088), and the New Century Excellent Talents Project of Minnan Normal University (No. MX14001).




\begin{thebibliography}{99}


\bibitem{Newman:2003}
M. E. J. Newman, The structure and function of complex networks. \emph{SIAM Rev.} {\bf 45}, 167 (2003).

\bibitem{BLMCH:2006}
S. Boccaletti, V. Latora, Y. Moreno, M. Chavez, and D. U. Hwang, Complex networks: Structure and dynamics. \emph{Phys. Rep.} {\bf 424}, 175 (2006).

\bibitem{BA:1999}
A.-L. Barab\'asi and R. Albert, Emergence of scaling in random networks. \emph{Science} {\bf 286}, 509 (1999).

\bibitem{JMBO:2001}
H. Jeong, S. P. Mason, A.-L. Barab\'asi, and Z. N. Oltvai, Lethality and centrality in protein networks. \emph{Nature} {\bf 411}, 41 (2001).

\bibitem{AB:2002}
R. Albert and A.-L. Barab\'asi, Statistical mechanics of complex networks. \emph{Rev. Mod. Phys.} {\bf 74}, 47 (2002).

\bibitem{SBGC:2007}
F. Sorrentino, M. Di Bernardo, F. Garofalo, and G. Chen, Controllability of complex networks via pinning. \emph{Phys. Rev. E} {\bf 75}, 046103 (2007).

\bibitem{RME:2009}
A. Rahmani, M. Ji, M. Mesbahi, and M. Egerstedt, Controllability of multi-agent systems from a graph-theoretic perspective. \emph{SIAM J. Contr. Optim.} {\bf 48},
162 (2009).


\bibitem{Liu2011}
Y. Y. Liu, J. J. Slotine, and A.-L. Barab\'asi, Controllability of complex networks. \emph{Nature} {\bf 473}, 167 (2011).

\bibitem{Yuan2013}
Z. Yuan, C. Zhao, Z. Di, W.-X. Wang, and Y.-C. Lai, Exact controllability of complex networks. \emph{Nat. Commun.} {\bf 4}, 2447 (2013).


\bibitem{LB:2016}
Y. Y.Liu and A. L. Barab\'asi, Control principles of complex systems. \emph{Rev. of Mod. Phys.} {\bf 88}, 035006 (2016).

\bibitem{NV:2012}
T. Nepusz and T. Vicsek, Controlling edge dynamics in complex networks. \emph{Nat. Phys.} {\bf 8}, 568 (2012).

\bibitem{YZD:2013}
Z. Yuan, C. Zhao, W.-X. Wang, Z. R. Di, and Y.-C. Lai, Exact controllability of multiplex networks. \emph{New J. of Phys.} {\bf 16}, 103036 (2014).

\bibitem{RR:2014}
J. Ruths and D. Ruths, Control profiles of complex networks. \emph{Science} {\bf 343}, 1373 (2014).

\bibitem{WNL:2012}
W.-X. Wang, X. Ni, Y.-C. Lai, and C. Grebogi, Optimizing controllability of complex networks by minimum structural perturbations. \emph{Phys. Rev. E} {\bf 85}, 026115 (2012).

\bibitem{XLHB:2014}
Y.-D. Xiao, S.-Y. Lao, L.-L. Hou, and L. Bai, Edge orientation for optimizing controllability of complex networks. \emph{Phys. Rev. E} {\bf 90}, 042804 (2014).

\bibitem{ZWL:2015}
C. Zhao, W. X. Wang, Y. Y. Liu, and J. J. Slotine, Intrinsic dynamics induce global symmetry in network controllability. \emph{Sci. Rep.} {\bf 5}, 8422 (2015).

\bibitem{ZZW:2017}
C. Zhao, A. Zeng, R. Jiang, Z. Yuan, and W. X. Wang, Controllability of flow-conservation networks. \emph{Phys. Rev. E} {\bf 96}, 012314 (2017).

\bibitem{GLDB:2014}
J. Gao, Y. Y. Liu, R. M. D\'Souza, and A. L. Barab\'asi, Target control of complex networks. \emph{Nat. Commun.} {\bf 5}, 5415 (2014).

\bibitem{IGS:2015}
F. L. Iudice, F. Garofalo, and F. Sorrentino, Structural permeability of complex networks to control signals. \emph{Nat. Commun.} {\bf 6}, 8349 (2015).

\bibitem{JLC:2013}
T. Jia, Y. Y. Liu, E. Cs\'oka, M. P\'osfai, J. J. Slotine, and A.-L. Barab\'asi, Emergence of bimodality in controlling complex networks. \emph{Nat. Commun.} {\bf 4}, 2002 (2013).

\bibitem{JP:2014}
T. Jia and M. P\'osfai, Connecting core percolation and controllability of complex networks. \emph{Sci. Rep.} {\bf 4}, 5379 (2014).


\bibitem{core_percolation1}
M. Bauer and O. Golinelli, Core percolation in random graphs: a critical phenomena analysis. \emph{Eur. Phys. J. B} {\bf 24}, 339 (2001).

\bibitem{core_percolation2}
M. Bauer and O. Golinelli, Exactly solvable model with two conductor-insulator transitions driven by impurities. \emph{Phys. Rev. Lett.} {\bf 86}, 2621 (2001).

\bibitem{core_percolation3}
Y. Y. Liu, E. Cs\'oka, H. Zhou, and M. P\'osfai, Core Percolation on Complex Networks. \emph{Phys. Rev. Lett.} {\bf 109}, 205703 (2012).

\bibitem{GKK2001}
K. I. Goh, B. Kahng, and D. Kim, Universal behavior of load distribution in scale-free networks. \emph{Phys. Rev. Lett.} {\bf 87}, 278701 (2001).

\bibitem{Dor2013}
N. Azimi-Tafreshi, S. N. Dorogovtsev, and J. F. F. Mendes, Core organization of directed complex networks.\emph{Phys. Rev. E} {\bf 87}, 032815 (2013).




\bibitem{data-air}
V. Colizza, R. Pastor-Satorras, and A. Vespignani, Reaction-Diffusion processes and meta-population models in Heterogeneous Networks. \emph{Nat. Phys.} {\bf 3}, 276 (2007).

\bibitem{data-int}
M. Faloutsos, P. Faloutsos, and C. Faloutsos, On power-law relationships of the Internet topology. \emph{ACM SIGCOMM} {\bf 29}, 251 (1999).

\bibitem{data-kdd2005}
J. Leskovec, J. Kleinberg, and C. Faloutsos. Graphs over time: densification laws, shrinking diameters and possible explanations. \emph{ACM SIGKDD}, 177 (2005).

\bibitem{data-ema}
J. Leskovec, K. Lang, A. Dasgupta, and M. Mahoney, Community Structure in Large Networks: Natural Cluster Sizes and the Absence of Large Well-Defined Clusters. \emph{Int. Math.} {\bf 6}, 29 (2009).

\end{thebibliography}
\end{document}